\documentclass[manuscript,screen,nonacm]{acmart}
\AtBeginDocument{%
  \providecommand\BibTeX{{%
    \normalfont B\kern-0.5em{\scshape i\kern-0.25em b}\kern-0.8em\TeX}}}

\usepackage{colortbl}
\begin{document}

\title{To Police or to Guide: How Higher Education Computer Science Instructors Design and Implement Generative AI Policies}


\author{Xingjian (Lance) Gu}
\affiliation{%
  \institution{University of Michigan School of Information}
  \city{Ann Arbor}
  \state{MI}
  \country{USA}
  }
\email{xjgu@umich.edu}

\author{Wells Lucas Santo}
\affiliation{%
  \institution{University of Michigan School of Information}
  \city{Ann Arbor}
  \state{MI}
  \country{USA}
  }
\email{ephemera@umich.edu}

\author{James M. Zumel Dumlao}
\affiliation{%
  \institution{University of Michigan School of Information}
  \city{Ann Arbor}
  \state{MI}
  \country{USA}
  }
\email{jamesmzd@umich.edu}

\author{Barbara Ericson}
\affiliation{%
  \institution{University of Michigan School of Information}
  \city{Ann Arbor}
  \state{MI}
  \country{USA}
  }
  
\email{barbarer@umich.edu}


\begin{abstract}
While generative AI tools are directly changing how undergraduate computer science is learned and taught, they are also reshaping the relationships between instructors and students. In contrast to existing tool-oriented research on how instructors view and adopt AI, this study investigates how instructors think about their roles and responsibilities to students through their course AI policies. Based on 13 semi-structured interviews with CS instructors in the US, we found that while instructors recognize that AI tools could harm student learning, AI policies primarily seek to AI-proof assessments without directly addressing student learning. Although policies such as switching to paper exams can preserve assessment integrity in the short term, instructors report extra burden of policing student AI use behaviors and worsening relationships with students. Based on the experiences of several interviewees, we make recommendations on AI policies that are more learning-oriented and could guide students toward healthier AI usage instead.
\end{abstract}



\maketitle

\section{Introduction}
Since generative AI (GenAI) tools became publicly available in 2022, they have moved from novel objects inspiring speculations to unremarkable parts of routine workflows. Undergraduate computer science (CS) students widely adopted GenAI tools when coding \cite{hou_evolving_2025}, many teaching assistants (TAs) began using AI for grading and lesson planning \cite{barkhuff_situated_2026}, and CS instructors are integrating GenAI-based teaching and learning tools into their courses \cite{lau_barriers_2026}. Within the research community, these trends have sparked numerous investigations and discussions about the relationships between students and AI as well as instructors and AI---how do students and instructors use AI \cite{manley_examining_2024, sevic_ai_2025}? Do they trust AI \cite{kann_students_2025, lyu_understanding_2025}? How can we make people more proficient with AI \cite{gu_ai_2025, wilton_where_2022}? 

While much of the attention is rightfully devoted toward the new technology that has the potential of altering the higher education landscape, it is important to recognize that the influence of technology is mediated by the social structures around it \cite{baym_personal_2010, padiyath_insights_2024}, and thus the research that focuses on AI tools cannot by itself tell us how CS education is changing. Teaching and learning CS still primarily take place in the interactions between instructors and students in formal schooling settings, so to understand shifts in CS education, we must also examine how instructor-student relationships are influenced by AI. To better understand the instructor-student relationships in the context of GenAI tools, CS instructors' course AI policies provide an ideal lens. By examining how instructors design and enforce AI policies in their classrooms, we can get insights into both instructors' observations and beliefs about how AI has affected their students, and what they think about their roles and responsibilities to students in light of AI. 

For the purpose of this study, AI policies are broadly defined as rules governing student's usage of AI tools, as well as related adjustments to pedagogy, assessment, grading, and academic integrity made in response to AI. While these policies are usually committed in writing in syllabuses, not all policies are always enforced as written, and not all policies are written down. As such, it is important not to confine the research to syllabus texts. As students' learning experiences are directly shaped by their instructors' policies \cite{luo_jess_how_2025, luo_jess_critical_2024}, understanding AI policies can shed light on how instructor-student relationships are reacting to GenAI tools. This study seeks to answer the following research questions:

\textbf{RQ1:} What do the instructors of undergraduate CS courses report as the effects of GenAI tools on teaching and learning?

\textbf{RQ2:} What AI policies have these instructors implemented, and why?

To address these research questions, we conducted 13 semi-structured interviews with current instructors of undergraduate CS courses across the US. The interviewees are from nine higher education institutions of different types, and they teach a diverse range of CS courses, ranging from CS1 to various advanced electives. The interviews cover their teaching background, what AI policies they have on paper, how they implement and enforce their policies in practice, and the reasoning behind their policies. We conducted reflexive thematic analysis to construct patterns of how instructors discuss their policies. 

Based on the interviews, we found that 1) instructors perceived two types of harms from AI---AI misuse can lead to learning harms to students, and also assessment harms to instructors by undermining their evaluation of students. 2) While some instructors are innovating on policies that can address learning harms to students, most instructors more readily reacted to assessment harms by AI-proofing their assessment and grading approaches, which can hinder student learning. 3) Due to the difficulty of policing AI use, instructors are placing responsibility of regulating student behavior on the student themselves.

These findings have important implications for CS education research. They show that \textbf{assessment-oriented AI policies only indirectly address AI use at best, while directly pressuring students to bear the negative consequences of AI misuse in hopes of students will change their learning strategy.} CS instructors should not expect students to have complete agency over their AI use, which is partially driven by the broader societal push for problematic GenAI tools \cite{flenady_cut_2026}. To better support student learning and maintain the integrity of assessments, instructors should consider shifting attention to policies that mitigate AI learning harms directly instead. Based on the instructor accounts, we put forth several of these learning-oriented AI policies that could be adopted without needing significant changes to the curriculum, including having more transparent communications about AI to students, modeling good AI use, and adopting more lower-stake formative assessments to facilitate metacognition.

\section{Prior Literature}
The impact GenAI has on CS education stems from its capabilities. Since their commercial launch, studies have shown GenAI tools are able to generate code that can pass CS assignments \cite{doughty_comparative_2024}, give easily comprehensible code explanations \cite{leinonen_comparing_2023, qiu_one_2026}, and provide debugging support \cite{macneil_decoding_2024, ma_how_2024}. Based on these capabilities, there is significant optimism that GenAI tools can augment CS teaching and learning, by providing more timely and targeted learning support to students, and support instructors with new tools for teaching and grading \cite{leinonen_using_2023, liu_teaching_2024, prather_robots_2023, rahman_chatgpt_2023, hou_codetailor_2024}. However, there are also concerns over AI disrupting CS education, such as cognitive offloading \cite{jose_cognitive_2025, prather_beyond_2025}. This section will review the field's current understanding of how AI influences CS teaching and learning, and highlight the needs for better understanding CS course AI policies and their effects.

\subsection{Effects of AI Use on Learning}
There is an increasing amount of empirical evidence on how using GenAI tools could influence learning. In contrast to research on how AI may increase productivity in workplaces \cite{noy_experimental_2023}, the effects of AI on learning is mixed. Prior findings show one way AI could undermine learning is via cognitive offloading \cite{gerlich_ai_2025}. Studies show that relegating cognitive load of programming tasks to AI can reduce code reading and debugging skills \cite{shen_how_2026}, reduce critical thinking skills \cite{gerlich_ai_2025}, reduce self-efficacy \cite{lee_relying_2026}, and lead to less perceived ownership of ideas and depth of thought \cite{baldeo_generative_2026}. There are also studies that suggest AI use could mislead students into think they have mastered coding skills when they have not, thus increasing the need for metacognition \cite{dawson_cognitive_2025, tankelevitch_metacognitive_2024, fan_beware_2024}. 

The availability of GenAI tools are also shifting social dynamics among students, as student help-seeking from instructors or other human sources has declined sharply, largely attributed to GenAI tool use \cite{hou_effects_2024, hou_evolving_2025, hou_all_2025}. AI can potentially assume the role of a participant in social learning activities \cite{tate_generative_2026}, but their effectiveness in those roles is questioned given their lack of a subjective self and lived experiences \cite{cambridge_theoretical_2024, flenady_cut_2026}.

While these insights help reveal the mechanisms through which AI use could influence learning, it is unclear how prevalent AI use is among CS students. Numerous studies attempted to gauge student AI usage and reported a wide range of results, from reporting relatively little use \cite{prasad_generating_2023} to widespread adoption \cite{manley_examining_2024}. This variability is unsurprising, because student adoption of AI tools depends on student perceptions and beliefs, which can vary widely across different learning contexts \cite{padiyath_insights_2024, yilmaz_augmented_2023, amoozadeh_trust_2023}. Adoption of GenAI tools among students has also increased over time \cite{hou_evolving_2025}. The influence of AI tools on a real world classroom is further complicated by the fact that they have heterogeneous effects based on individual student characteristics, helping students who are ahead and impairing the learning of students who are struggling \cite{prather_widening_2024}. As such, instructors play an important role in guiding students towards more beneficial uses of AI and in mitigating the harms of misusing AI \cite{pallant_mastering_2026}.

\subsection{Instructor Reactions to AI}

For instructors, the trouble brought by GenAI tools goes beyond impairing student learning. GenAI tools also undermine the integrity of assessments, as models since 2022 have been demonstrated to perform well on many programming assignments and tests \cite{finnie-ansley_robots_2022, finnie-ansley_my_2023, doughty_comparative_2024, joshi_chatgpt_2024, hou_more_2024}, and their capabilities have only increased since then. Programming education as a field is more susceptible to GenAI disruptions, due to AI tools being capable of a wide range of tasks that students are supposed to learn \cite{rebelsky_talking_2026}. These led to increased concern from instructors over students using AI to cheat \cite{cotton_chatting_2024, kizilcec_perceived_2024, harrington_did_2025}.

Given the challenges posed by AI, it is important to know how instructors react to them, as their actions shape student learning experiences. Prior studies have explored CS instructors' attitudes towards GenAI, which generally involve a mixture of anxiety and optimism over how AI could both harm and support learning, as well as different levels of familiarity with AI tools and trust in them \cite{sheard_instructor_2024, lyu_understanding_2025}. Lau and Guo found in an interview study in 2023 that after CS instructors witnessed the capabilities of GenAI tools, they envisioned both short-term course adjustments and long-term curricular changes to adapt to AI \cite{lau_ban_2023}. In 2026, Lau and collaborators found many CS instructors are open to integrating AI tools into their teaching, but face barriers in practice, such as lack of time and resources and push back from colleagues \cite{lau_barriers_2026}. These studies' interview topics are speculative and focus on instructor reactions to the technology of GenAI itself. 

Nevertheless, students have been using AI tools regardless of instructors attitudes toward AI. Thus, these studies leave gaps in understanding how instructors are managing and reacting to student AI usage since 2022, and how the instructor-student relationship has been affected by AI. Understanding the latter is especially important against the backdrop of GenAI tools, since students still rely on instructors for most of their learning in higher education institutions. By changing how students could fulfill their schoolwork, AI is directly altering the nature of instructor-student interactions \cite{kahn_teacher_2025, luo_jess_how_2025}. For example, Luo found that college students perceive a power imbalance between themselves and instructors due to the lack of transparency in AI policies and their enforcements, and a general erosion of trust between students and instructors due to AI \cite{luo_jess_how_2025}. While there are many studies that investigate student-AI interactions \cite{manley_examining_2024, boguslawski_programming_2024, kann_students_2025, harrington_did_2025, hou_evolving_2025} and instructor-AI interactions \cite{lau_barriers_2026, sevic_ai_2025, lau_ban_2023, kizilcec_perceived_2024, lyu_understanding_2025}, we should not lose sight that instructor-student relationships are still crucial and worth examining. This study seeks to contribute to the understanding of how instructors view their relationships and responsibilities with students, in the wake of GenAI.

There are also efforts at analyzing the AI policy language instructors include in syllabuses and how cheating is redefined to include AI misuse. From instructor's perspectives, certain AI uses are often considered cheating, such as submitting AI-generated code as one's own work \cite{cotton_chatting_2024}. However, not all AI uses fall under cheating, and past research on CS instructor's efforts on deterring cheating shows that it has mixed results at improving student learning \cite{albluwi_plagiarism_2019}, so understanding AI policy through the lens of cheating would be too narrow. In terms of written policies, institutions recognize the importance of having AI policies \cite{an_investigating_2025, luo_jess_critical_2024}, but the actual implementation at the course level can vary drastically, ranging from banning AI completely to having no restrictions \cite{ali_analysis_2025, bui_adoption_2026, tong_what_2025}. This variability and often lack of transparent communication about AI policies causes confusion among students, as they are not certain what uses of AI are allowed and what they can disclose to instructors and teaching assistants (TAs) without risking penalty \cite{corbin_wheres_2025, dewan_engineering_2025, luo_jess_how_2025}. Analyzing AI policies as written on paper is also not necessarily indicative of how they are enforced in practice, or whether they are effective. This study aims to fill those gaps by exploring how instructors both design and enforce AI policies, and how they view the outcome of their policies. 

\section{Methods}

\subsection{Data Collection}
To address the research questions, we conducted 13 hour-long semi-structured interviews with instructors who were actively teaching computer science courses in higher education institutions. The interview study was reviewed and granted exemption by the IRB. The interviews were conducted remotely over Zoom, and each interviewee was compensated with \$50 for participating. Most interviewees were recruited by purposive sampling, with some instructors recruited by snowball sampling as we encouraged interviewees to forward the recruitment message to others who are qualified and interested in participating. 

To fully understand each instructor's approach to designing, implementing, and enforcing their AI policies, the interview protocol included four sections. It began with a warm-up section to understand the instructor's teaching background and course context, such as what courses they have taught over the last three years. The protocol then explored their AI policy on paper, as in what do they officially include in their printed syllabuses. It was followed by questions about the AI policy enforcement, to see what evidence instructors use to determine if their students are following their policies and what they did if they saw violations. Lastly, it asksed about AI policy design, such as what factors motivated their policies. Sample questions from each section are shown in Table \ref{table:interview_protocol}. In the cases where an instructor teaches multiple courses that each have different AI policies, they are asked to answer the questions for each of those courses if time permitted.

\begin{table}[ht]
\begin{tabular}{lll}
\hline
        \textbf{Interview Section} & \textbf{Example Question}\\
        \midrule
       Teaching Context  & What are the learning objectives of your course? \smallskip \\
       
       AI Policy on Paper  & What is your policy on generative AI use in the syllabus? \smallskip \\

       AI Policy Enforcement  & What do you do if students violate your AI policy? \smallskip \\

       AI Policy Design  & How does your teaching philosophy shape your AI policies? \smallskip \\
       \bottomrule
    \end{tabular}

    \caption{Example questions from the interview protocol}
    \label{table:interview_protocol}
\end{table}

\subsection{Interview Participants}

The interviewees come from diverse teaching contexts, including nine higher education institutions located in the US, and have taught a wide variety of CS courses. Although a slim majority of the interviewees have recently taught CS1 courses, all interviewees also had recent experience teaching more advanced elective courses. Several interviewees have also taught courses designed for non-CS major students. Thus, these interviews can provide us with a more complete picture of AI policies across different contexts, given that previous efforts at understanding the effect of GenAI on CS education have primarily focused on CS1. Since the population of higher education CS instructors in the US is relatively small, we opted to take extra care in protecting the participants' identities and not report their demographic information. The class sizes for each instructor's courses are based on the largest course they teach, with small signifying classes with fewer than 30 students, medium from 31 to 100, and large for classes with over 100 students. More detailed information of the interviewees' teaching contexts is in Table \ref{table:teaching_context}.

\begin{table}[ht]
\begin{tabular}{llll}
\hline
ID  & Institution Type            & Recently Taught Courses & Class Size                         \\ \hline
I1  & Public Research University  & Data Structures, Data Science  & Medium                   \\
I2  & Public Research University  & CS1, CS2, Computer Systems, Data Structures  & Large     \\
I3  & Public Research University  & System Administration, Game Programming   & Medium       \\
I4  & Private Research University & CS1, Human-Computer Interaction, CS Capstone   & Medium  \\
I5  & Public Research University  & CS1, Computer Systems, Computer Architecture  & Large   \\
I6  & Private Research University & CS1, Data Structures  & Small                            \\
I7  & Private Research University & CS1, Automata Theory  & Large                           \\
I8  & Public Research University  & Data Science   & Medium                                  \\
I9  & Public Research University  & CS2, Problem Solving   & Large                          \\
I10 & Private University          & Software Engineering, Human-Computer Interaction  & Small\\
I11 & Private Research University & CS1, Game Design    & Large                             \\
I12 & Private Research University & Artificial Intelligence, Data Science, CS Education     & Medium       \\
I13 & Public Research University  & CS1, CS2, Artificial Intelligence      & Large          \\ \hline
\end{tabular}

\caption{Teaching Context of the Interviewees}
    \label{table:teaching_context}
\end{table}

\subsection{Data Analysis}
To analyze the data, we applied reflexive thematic analysis to identify patterns of how instructors design and implement their policies \cite{braun_using_2006}. As explained by Braun and Clarke, reflexive thematic analysis is an interpretivist method that acknowledges how the meaning of interviewees' words is constructed by the researchers, instead of objectively representing the interviewees' interior thoughts and can `emerge' for researchers to find \cite{braun_reflecting_2019, braun_toward_2023}. When applying this method, researchers should recognize their active and reflexive role in analyzing data, instead of obscuring their subjectivity with more positivist metrics, such as computing inter-rater reliability \cite{braun_toward_2023}. By extension, the findings from the analysis are not meant to be representative of a broader population. Rather, they serve as an exploratory source for theorizing and informing future quantitative studies. As such, we will document our analysis process in detail in lieu of reporting inter-rater reliability. 

Following the six phases of thematic analysis \cite{braun_using_2006}, the three researchers began by closely reading through each transcript and taking notes, paying special attention at the following: intentions and theories behind policies, tensions between policy language and enforcement, and observations of student behaviors. We met over several meetings to compare our notes and discuss our interpretations. During these meetings, we shared our initial codes, such as `time and human cost of detecting AI use', `focus on debugging skills', and `switch to pen and paper exam', and theorized overarching patterns of meaning that could explain and motivate seemingly different policies. These were refined into thematic codes, such as `second-order harms from assessments', `fraying student-instructor relationships', and `responsibility rhetoric'. We then extracted representative quotes from the transcripts. We compiled the quotes, reviewed the quotes together, and renamed and consolidated themes where necessary, before producing the analysis. As part of the reflexive methodology, we the researchers want to acknowledge our positionality. All three researchers have recent programming teaching experience, with two of the three having taught computer science prior to the advent of GenAI. The researchers have varying degrees of skill and trust in using GenAI tools, ranging from daily use to non-use due to ethical concerns. All three researchers are motivated by interest in AI policies and their effects.

\section{Analysis}
\subsection{AI's Learning Harms and Assessment Harms}
To understand the instructors' decisions when creating their AI policies, it is important to first recognize what issues they observe and how they believe these issues were caused by AI. This immediately poses a major challenge. Despite GenAI tools' accessibility over the last three years and general anecdotal agreement that AI is impacting learning, instructors find their actual effects on students complex and multi-faceted. 

\subsubsection{Students Are Shortcutting Their Learning}
The most commonly shared concern among instructors with student AI usage is the risk of cognitive offloading, or `shortcutting their learning,' as I9 and I13 phrased it. Cognitive offloading and its harmful effects to learning have been observed in empirical studies \cite{baldeo_generative_2026, gerlich_ai_2025}, and instructors have observed students in their own classrooms with similar behaviors:
\begin{quote}
    \textit{When students hit a problem, like a compile error or something, and they just immediately go to AI for the answer. Rather than first trying to debug it themselves\ldots They couldn't struggle as much cognitively through difficult problems, and that's been probably detrimental to their learning.} (I4)
\end{quote}
Instructors found learning harms due to cognitive offloading especially acute when students are still forming fundamental skills, such as learning basic syntax and debugging. Several instructors who teach CS1 courses, including I5, I6, and I11, expressed that they believe students should not use AI tools in their CS1 courses due to the necessity of deliberately practicing basic coding skills:
\begin{quote}
    \textit{In intro, they are truly developing their debugging muscles, their coding muscles. I am much more strict in intro, because that is them having to learn, if they are gonna have to use AI in the future, they're gonna have to be able to debug.} (I5)
\end{quote}
Instructors' worry over AI shortcutting learning and undermining basic skill formation is certainly not new, as it was one of the potential concerns speculated by instructors in Lau and Guo's study in early 2023 \cite{lau_ban_2023}. However, our evidence also suggests the impact is not contained within introductory courses. When students who have not mastered fundamental skills still managed to pass the CS1 courses due to AI, it creates problems for instructors of CS2 and beyond:
\begin{quote}
    \textit{I still have some students who I think, honestly, cheated their way through previous courses, and they're, in the nicest way\ldots they don't know anything. Like, some of them can't really do variable assignments sometimes, and I'm like, `}\textbf{I don't understand how you've made it into this course}\textit{,' and then I think they just feel really stuck, and they end up cheating more.} (I6, emphasis ours)
\end{quote}
Even for students who have acquired fundamental skills and are taking more advanced elective courses, instructors have nevertheless noticed detrimental effects from cognitive offloading. From the observation of I12, students working on an open-ended data visualization demonstrated less creativity since the advent of GenAI tools, turning in projects that were \textit{`super boring. Like, who cares? I guess because LLMs are boring.'} These instructor observations suggest that past speculations about AI's learning harms to skill formation materalized, but the harms are not confined to the fundamental skills. Reduced cognitive engagement harms learning across course contexts.

\subsubsection{Illusion of Competence}
Compounding with the issue of cognitive offloading, as many instructors noticed, students are not always metacognitively aware that they might be learning less with AI. Several instructors (I3, I4, I6, I11, I13) remarked that some students have an illusion of competence, believing that they have mastered the content when they have not:
\begin{quote}
    \textit{One of the dangers of AI is that \textbf{it fools students into thinking they know more than they do}. They have this cognitive dissonance between what they know and what they think they know. And I think we're seeing the results of that now in our courses.} (I4, emphasis ours)
\end{quote}
While there were other ways where students could cognitive offload their learning before GenAI tools were available, such as by plagiarizing other's work, instructors observe that AI distorts student's metacognition in a new way. Tools such as Copilot frame themselves as helping tools that merely auto-completes codes that students have learned. Instructors notice that this causes students to erroneously believe that if they can make sense of code generated by AI, they have learned that concept, without realizing that cognitively offloading the work of writing the code is harmful to learning:
\begin{quote}
    \textit{I think a student could do that with very good intentions, and yet still inadvertently circumvent the learning process, because maybe I want them to have that productive struggle of trying to debug the code themselves. Because it's always harder to work something out ourselves than it is to hear an explanation and say, }`oh yeah, I got it.'\textit{ I tell my students, if all it took was an explanation, I would just tell you, and then we'd be done.} (I13)
\end{quote}
As I13 pointed out, instructors can intervene to dispel this conflation of familiarity with mastery and of code tracing for code writing, and the resulting illusion of competence. This is especially necessary, given how easy it is for students to `inadvertently circumvent learning' now:
\begin{quote}
     \textit{Because Google Colab, they also have the AI plugin, right? So, I have to tell the students,} `Okay, let's get to the settings, and then turn the AI assistance off.’ \textit{Because otherwise, the student doesn't really need to do anything. They just, the app, like, you know, like, automatically prompts the answer for them to type. }(I3)
\end{quote}
In a sense, students are being actively misled by the promises of GenAI tools as productivity boosters, as I6 puts it: \textit{`[Students]'re like, oh but people in industry use it, and I'm like, yeah, but they know what they're doing. Like, you're learning. You're learning how to do any of these things at the basic level. You're not trying to use it as a tool to make your work faster.'} In that sense, the instructors are not only managing the effects of GenAI on students, but also the marketing and society-wide discussions surrounding it. While the design of a GenAI tool might not inherently induce an illusion of competence, the instructors found it necessary to also directly address and push back on students' expectations surrounding the tools.

\subsubsection{Social Isolation}
Beyond the cognitivist concerns with how AI influences individual students' learning, most instructors also remarked on changes in social learning activities as well, such as office hours attendance, posting on help forums, and teamwork. GenAI tools potentially disrupting social interactions among students has been reported \cite{hou_all_2025}, but the reduced interaction between instructors and students adds a new dimension to the social isolation. To instructors, the most directly observable trend is less office hours traffic (I1, I6, I7, I8, I9, I11):
\begin{quote}
    \textit{The trends are definitely reduced frequency of help resources within the course, specifically those attached to humans. Posted questions are down---the frequency, the quantity of questions. Office hours traffic is down. And students coming to office hours with code they cannot explain is up.} (I11)
\end{quote}
It is worth noting in I11's account that not only was there a quantitative decrease in student seeking help from human sources, there was a qualitative shift in the nature of help-seeking. The \textit{`code they cannot explain'} brought to I11 by students is most likely attributed to students generating code with AI, which then hints at a potential reason instructors suspect why help-seeking from humans is decreasing:
\begin{quote}
   \textit{[AI] does deprive [students] of an opportunity for learning. If I was turning in homework and someone else was doing it the entire time, then if something goes wrong with that homework, I don't know how to go fix that. And if I take it to office hours with a TA, I don't even know what questions to ask.} (I8)
\end{quote}
Based on I8's perspective, the reduced office hours volume is not just the result of direct displacement (where students used to ask a question to TAs or instructors, now they ask a chatbot). Rather, there are new social dynamics at play, where some students feel shame about their AI use, and consequently withdraw from seeking human help due to perceived risks of being found out. This AI stigma has been reported in prior studies \cite{bao_ai_2025, hou_all_2025, luo_jess_how_2025}. This new dynamic puts additional strain on how instructors need to maintain their relationships with students.

Instructors have also reported seeing less interactions among students. I10 observed that GenAI use can cause frictions between students when they are assigned to do teamwork: \textit{`I think I've had two cases where the rest of the team in the teamwork evaluations said, like ``they produced all this code, but it didn't fit with what we were doing, and we had to rewrite it all.''} Yet beyond these structured interactive learning activities, some instructors have noticed less social interactions between students in general:
\begin{quote}
    \textit{I see them much more independent, but not in a good way, as in solitary. At least before, when you cheated, you have to make friends. They don't even make friends. It sounds horrible, but that's how I see it. They are very much solitary. They seem more depressed, they seem more down.} (I5)
\end{quote}
While it is difficult to directly attribute I5's observed solitude among students to GenAI usage, it nonetheless connects with both prior findings on erosion of social interactions due to AI \cite{hou_all_2025}, as well as the downward trend in seeking help from instructors. Given the importance of social interactions to learning \cite{vygotsky_mind_1978}, and that most AI tools available to students lack the capability to fully participate in social learning as a study partner \cite{cambridge_theoretical_2024}, the social isolation of students observed by instructors could constitute a very real learning harm.

\subsubsection{Scores not Reflective of Learning}
Beyond the three ways instructors identified where AI use could harm learning at the level of individual students---cognitive offloading, illusion of competence, and social isolation---instructors are also concerned about how AI impacts their assessments. Assessment harm is distinct from learning harms to instructors. While only some students' learning is undermined due to AI, instructors can no longer tell who among their students have truly learned and who need more help based on their submitted work. This can lead to generalized distrust toward assessments from instructors:
\begin{quote}
    \textit{I used to have a reasonable idea that if they completed those assignments, then they understood programming. It was clear from people who'd done the homework, they were mostly doing well on the exams. That was fine. But now I have no reasonable way to know whether they actually struggled through them, or if they just generated them with AI, so \textbf{I can't really give out course credit for that anymore.}} (I4, emphasis ours)
\end{quote}
Even though instructors often refer to student cheating in the context of AI misuse, we avoid using the word `cheating' when discussing assessment harms, because it encompasses many violations of academic integrity, the exact rules on which can vary across institutions. `Cheating' also focuses on the actions of students, while we focus on the effects on instructors. As I4 describes, a critical issue perceived by instructors is not necessarily that students are cheating, which have always happened before AI too, but rather that AI is making it more difficult than before to tell apart work submitted by students who did it without external help from work generated by AI. It is both because AI is more readily available than traditional means of cheating, as noted by I2: \textit{``LLMs have somewhat removed [the friction] from people who are choosing to do this, right? It used to be you had to pay Course Hero, or upload enough documents to Course Hero to get something, or pay Chegg for a selected answer\ldots it's just getting easier and easier to skip the step of doing it yourself,''} but also because AI-generated output is not deterministic, which circumvents some of the traditional methods of detecting plagiarism.

Even though it is hard to have concrete evidence on whether a student has used AI for their work, instructors do see new and unusual patterns that suggest AI misuse. For instance, since the introduction of GenAI, instructors see more students who turn in perfect homework but perform poorly on exams: \textit{``Now, suddenly, I have students that are getting 10\%. Like, how do you get 10\% on a test when you're getting 100\% on assignments?''} (I6), or not knowing very basic concepts late in the semester: \textit{``Someone who's a pretty good student, but they couldn't remember if an array in Java started at 1 or 0, and I'm like, ‘If you're doing your own work all the time, you know it starts at zero.’ ''} (I9). These phenomena led instructors to recognize that they can no longer tell \textit{whether} their students are learning, which is a matter of assessment, intertwined but distinct from making sure their students are learning, which is a matter of teaching and learning. I1 summarized this dual effects of learning harm to students and assessment harm to instructors:
\begin{quote}
    \textit{The way I've looked at it nowadays is if you want people to think, you have to do it in class. Otherwise, anything else you're asking students to do, you have to question whether they're doing any thinking about it, which kind of sucks.} (I1)
\end{quote}
In aggregate, instructors perceive challenges from AI from two dimensions: AI is undermining student learning via multiple mechanisms, each requiring interventions, and AI is compromising their ability to assess student learning as a whole, as it becomes harder to tell who really mastered the content versus who just submitted AI-generated work. The AI policy items they adopt can thus be sorted based on whether the policy seeks to influence learning or assessment.

\subsection{Policing AI Use Incurs Second-Order Costs}
Even though instructors identified more ways AI might harm student learning than their assessments, because the assessments are more directly under their control, AI policies shared by instructors usually focused on assessments. 

\subsubsection{Detecting Signs of AI Use}
Many instructors mentioned that they have attempted to detect AI use in student submitted work, but the non-deterministic nature of AI-generated code makes traditional approaches, such as similarity scans, less effective. A feature of AI-generated code that instructors could still rely on for detection is that AI can output code with features not yet covered in class:
\begin{quote}
    \textit{the TAs see a lot of examples of truly human-written code, especially before AI was out there. And then they see examples of AI. AI does specific things, it uses features in Java that we do not allow. So, as soon as they see forbidden features, it's like, oh, why are they using this? They shouldn't even be using this in the course.} (I5)
\end{quote}
While some instructors use uncovered code, in combination with traditional similarity scanning tools to identify students who misuse AI (I2, I5, I6, I8), this policy can be taken further and become a formalized policy:
\begin{quote}
    \textit{We've ultimately arrived at this policy where we just have a very strict rule of you can only use things we learned in class\ldots On the first homework assignment last time, half the class got really marked down because they were using this extra function that we didn't [teach]. I kind of left that in as a honeypot, so maybe half the class was still using Gen AI.} (I1)
\end{quote}
Based on the experience of I1, this approach was certainly effective at detecting students who might have used AI, but it has clear limitations. For one, it would not detect students who used AI but were careful to remove disallowed code, especially if the policy is made known to students. If the policy is enforced strictly, it could also lead to false positives, where students who genuinely learned the content from other sources would be penalized. In other words, it forecloses learning that is not designed by the instructor. Lastly, even among instructors committed to detecting AI in student submissions, some acknowledge the human labor and time cost of detection work (I5, I6):
\begin{quote}
    \textit{I could have spent hours a day, every day, just doing reports, but I had no time. So, amongst that, I was only able to get to 25\% of reported cases to submit to student conduct\ldots All this is very human labor intensive. It is line by line, then looking at the replay. Sometimes the replay can be hours and hours of their work, and I would have to step through it all.} (I5)
\end{quote}

\subsubsection{Paper Exams is the Only Choice}
Given the difficulty of detecting AI in submitted work, many instructors opted to prevent the use of AI in their assessments instead. The most popular approach to achieve that is to use in-class, proctored exams. Furthermore, instructors would go further and make the exams pen-and-paper because, according to I5, \textit{`We did find cases of students using ChatGPT and online resources that they weren't allowed to use on their laptops during the quizzes, even though they were proctored by their TAs.'} The practice of switching to paper exam is surprisingly common, with 9 interviewees adopting some form of it (I1, I5, I6, I7, I9, I10, I11, I12, I13), with I4 still allowing laptops but moving assessments in-class. Resonating with the threat of AI on assessment integrity to instructors, the most cited reason for adopting paper exams is for instructors to ``truly'' know what their students have learned:
\begin{quote}
    \textit{For the summative assessments, those are going to be two midterms and a final. On paper, no technology. I strip away everything. \textbf{It's so far the only thing I've managed to do to truly understand how much did the student master} versus how much are they able to do with help of any sort, including AI.} (I11, emphasis ours)
\end{quote}
It is worth emphasizing again that moving to paper exams does not directly intervene on AI's learning harms. Giving students paper exams, as the instructors suggest, restore their confidence in the assessment by making AI use much more difficult. However, as some instructors simultaneously point out, paper exams do not necessarily help students learn better---it might actually cause some students to perform worse:
\begin{quote}
    \textit{Because I have test anxiety, I do not like a testing environment. I get stressed, you know. During an exam, suddenly 1 plus 1 equals 3 in my head. It is not fun for me, and I do not like it for my students. However, I see no choice but to do that, because I cannot find a way right now to evaluate the learning objectives without this closed book environment to see and evaluate what they have learned.} (I5)
\end{quote}
To I5, the adoption of paper exams is a matter of tradeoff. Despite understanding the problem with high-stake exams given the lived experience of test anxiety, I5 viewed it as a necessary cost to bear for the benefit of more reliably assessing the learning of students. However, as other instructors point out, paper exams do not reliably measure student learning without having the curriculum designed around it:
\begin{quote}
    \textit{I think the main thing is I don't have time to do backward design for [paper exams]. Because if I'm expecting students to do pen and paper on the final exam, then I have to give them those 3 to 4 or 5 opportunities to do practice on paper as well, throughout the term.} (I1)
\end{quote}
As I1 pointed out, switching to paper exams as an emergency measure might cause the summative assessment to become disjointed from the learning students have done throughout the semester. If instructors switch to paper exams based on the assumption that it can help expose students who are over-reliant on AI, but students might struggle with paper exams for other reasons like test anxiety, paper exams could not restore the integrity of assessment as the adopters would hope. For paper exams to truly work, I11 suggests, they need to be tailored to the learning objectives: \textit{`The reason we [provide a reference sheet] is because we're not assessing how well they can memorize specific things. It's more so about, how good are they with problem solving and computational thinking.'} 

\subsubsection{Make Passing Exams Mandatory}
In addition to adopting exam formats that instructors believe to be more reliable, many instructors also weigh the exams more heavily (I2, I4, I5, I6, I7, I9), sometimes to the extent that passing the exam is functionally required to pass the class. These adjustments range from shifting 10\% of total grades to exams, to eliminating grades from take-home assignments completely and putting all grades on proctored quizzes and exams. Making passing exams mandatory can also be formalized as a policy:
\begin{quote}
    \textit{We have a rule now, where if you don't get above a 50\% on all of your combined exams, you don't pass the course. Yeah, because there's just been problems where students really are bombing the tests, but because other things have a high enough weight, some of them occasionally can get close to having a passing score when they really shouldn't. Because even if we can't catch them cheating, it's like, how are you getting such a low score on these things, and then 100\%s on everything else?} (I6)
\end{quote}
The policy of adjusting grading weights is further evidence that AI learning harms and assessment harms are distinct, and an AI policy can be targeting only the assessment harm. As I6 explains, making passing exams necessary is driven by students who are believed to be cheating, but cannot be caught normally given the challenges posed by AI tools. In other words, the goal is to fail the students who likely misused AI to preserve the integrity of the course, as opposed to mitigate the learning harms caused by AI to those students. Some instructors (I2, I12) justify the policy changes by pointing out that high-stake exams can motivate students to take learning more seriously:
\begin{quote}
    \textit{I'm not going to be able to stop them from accessing an LLM on their own time. But, what I can do is make the reward for doing so minimal if you're doing it the wrong way, and maximize the reward if you're doing it the right way. So when I taught data structures last time, programming assignments were 30\% of the grade, there was reading for another 5\%, and the other 65\% came from exams.} (I2)
\end{quote}
Even then, I2's reasoning yields an insight that we will explore further later, which is that what some instructors assume is the reward, and thus would incentivize students, is the \textbf{grade}. As such, the underlying reasoning surrounding this AI policy is still a matter of assessment, not learning---make exams AI-proof so students would drop AI and because they want exam grades. However, there is no guarantee that students will adopt better learning strategies, especially if they are not aware of them.

\subsection{Guiding AI Use Does not Require Fundamental Changes}
In contrast to making assessment-oriented AI policies, making policy changes that can address student learning seem more daunting at first. Several instructors expressed their belief that they might need to fundamentally revamp their learning objectives and curriculum to adapt to AI: 
\begin{quote}
    Should [students] follow exactly the same path as people did 10 years ago? Probably not exactly the same path. But should they hardly learn how to code at all?\ldots 10 years from now, there will be some things that we will need to teach that are not on people's radar right now. Something about GenAI, I don't know, orchestration of agents, something like that, you know? (I13) 
\end{quote} 
However, several instructors have identified strategies to adapting their AI course policies that required neither making assessments more strict, nor fundamentally changing their courses.

\subsubsection{Transparent AI Instruction}
A few prior studies have explored how students react to course AI policies, and a recurring finding is that students highly value transparency \cite{adnin_examining_2025, sumilong_instructional_2025, luo_jess_how_2025}. Since AI policies can vary significantly between courses, students grow wary of admitting to using AI due to uncertainty over whether their uses might be disallowed \cite{luo_jess_how_2025, bao_ai_2025}. A simple yet effective intervention is to be transparent and constructive when discussing AI with students. 

To begin with, many instructors recognized that there exists a wide range of ways to use of AI, with some being helpful to learning while others are detrimental. As such, the question surrounding AI policy is not a binary question of whether to completely ban AI or not, but a gradient of what uses of AI are deemed helpful, and how to direct students to use AI in helpful ways: \textit{`I think there's a right and a wrong way to use Gen AI in education, and it would be a lot to ask students to have that discernment.'} (I13) This discernment is often baked into the a common AI guideline, namely students are allowed to consult AI, but not allowed to copy-paste generated code directly (I3, I5, I7, I9, I10). However, several instructors (I2, I4, I11, I12) make their intentions more explicit and model good AI uses directly in class:
\begin{quote}
    \textit{I do teach them how I think they should use AI in CS1. For example, I give them an example prompt where you give the AI your code, you give the AI your error message, and you tell the AI, don't tell me what's wrong with my code. Help me, ask me questions so I can think through how I should approach debugging this problem.} (I11)
\end{quote}
While instructors acknowledge that not all students will follow these suggestions, an approach can still be effective even if it does not convince everyone. As I11 points out, openly acknowledging and teaching students about AI can shift the classroom culture surrounding AI use: \textit{``Once I talked about it in lecture, and kept emphasizing that using AI is not the problem, no, learning is the problem. Then they're much more open and willing to confess to using it.''} Ultimately, if we recognize that students may be using GenAI tools regardless of whether they are banned in specific courses, it is better to direct students towards AI uses that are more conducive to learning.

Taking transparency on AI a step further, some instructors (I4, I6, I9, I12) directly incorporate AI literacy instructions into their courses, in the hopes that equipped with more knowledge about how AI works and how it influences society at large, students will be more capable at choosing constructive uses for themselves.
\begin{quote}
    \textit{We teach them about what large language models are, and what's behind today's AI. And then, as we teach across the different areas of computer science. We give them practice or information about how large language models might be used in those areas, and try to highlight risks and benefits, and have open dialogues.} (I9)
\end{quote}
AI literacy encompasses many distinct concepts, including how AI works under the hood, how to use AI tools effectively, and sociocultural influences of AI \cite{long_what_2020, gu_ai_2025}. In I9's example, the AI literacy instruction included many of these ideas, while others might focus more the technical aspects (I9, I12) or sociocultural aspects (I4) specifically. 

To take one final step in transparently communicating how AI can fit into the curriculum without an overhaul, instructors could consider how AI interacts with each of their learning objectives, and provide clear guidance on how students should use AI for each activity (I4, I6, I10):
\begin{quote}
    \textit{I think [the AI policy] depends on the goal of the assignment a little bit as well. So for the individual project in CS3, it is supposed to help you figure out where you are in terms of proficiency in Java. It's not very helpful if it's entirely AI-generated. On the other hand, the second project in the course, it's a group project, it mirrors maybe more software engineering practices, as I imagine they're going on in the workplace. And so I was more open to [AI].} (I10)
\end{quote}
This approach, while providing more granular control, also requires more direct communication with students as changing AI rules throughout a semester can be confusing. Hence, instructors adopting this approach would need to diligently explain to students what their expectations with AI are and why that is. However, as we will explore next, explicitly talking about learning objectives and how AI relates to them with students can be beneficial to their metacognition as well.

\subsubsection{Promote Metacognition with Formative Assessments}
Given the instructors' perception that a lack of awareness in some students that AI misuse is hurting their learning, many also made policies aimed at facilitating student metacognition. The most direct instantiation of this approach is to give students self-reflection forms after each assignment (I7, I11):
\begin{quote}
    \textit{This is called a reflect form, they have to fill out with every assignment. We have all these questions we ask them, like, }`It's fine to get help, just tell us who you got help from.' (I7)
\end{quote}
There is some implementation limitations to distributing self-reflection forms, as it adds additional artifacts for instructors to evaluate, which I7 also acknowledged: \textit{``Now, I haven't gone back and looked to see how many people actually use the ChatGPT website or something\ldots because we never look at those reflectors, unless we have to go back and look for something like when someone cheats.''} One could argue that these forms do not need to be graded, as having students perform self-reflection and leaving a paper trail is enough. However, there are findings that indicate students might take AI disclosure and self-reflection less seriously if they realize that have no stake and are not reviewed \cite{gonsalves_addressing_2025}. 

As such, a common alternative approach to promote metacognition is to include more frequent, formative assessments into the syllabus. This policy stems from the same logic behind assessment-oriented policies, where instructors require students to engage their skills independent of AI tools during in-class assessments. However, in contrast to making in-class exams into high-stake summative assessments for the main purpose of increasing assessment integrity for the sake of instructors, instructors who made them low-stake formative assessments used it to help make students realize AI's learning harms to them earlier:
\begin{quote}
    \textit{If they shortcutted their programming assignments, and then they took the first test, and they struggled with it, then they have a week. They can come see us to look at the questions. That is good for getting them to have human contact. And then they can try again, so they don't bury themselves. Like, if they over-rely on the LLM, they could still recover.} (I9)
\end{quote}
Among the interviewees, formative assessments took many forms. It could be a low-stake in-class exam (I1, I4, I8, I9) or an open-ended project where students are given feedback at multiple points (I6, I7, I10). In either case, it allows students who might have an inaccurate self-evaluation of their learning to fail in a low-stake scenario, so that they can adjust their learning strategies or have instructors provide more prompt interventions. In the specific case of I9, the formative assessments were directly connected to human help resources, which can also address the social isolation issue and further reduce reliance on AI.

\subsubsection{Inspire Motivation and Interest}
Lastly, some instructors view student's reliance on AI as a symptom of lacking interest or motivation in the subject, rather than a problem in and of itself. As I8 put it: \textit{`If there are a lot of people whose solutions look really similar to AI for a particular problem, that signals to me more that there might be an issue with that particular problem, and not the fact that students are using AI.'} Following this line of thinking, a few instructors believe that increasing student motivation in learning CS can help reduce their AI use (I6, I8, I9, I10):
\begin{quote}
    \textit{I form teams [by having students] sort of brainstorm potential areas of interest, topics, and then form teams around what they're interested in, so that, you know, I think it helps with motivation and interest, if people work on things that they actually care about.} (I10)
\end{quote}
Instructors who adopt this approach tended to view teaching and learning from a more social learning perspective, and would explicitly discuss how building a healthier learning environment can lead to the side product of students relying less on AI: \textit{``It's the course culture---not fighting against the students, but working with them to develop their own mental model and approach, creating good work, and having ownership of it, and learning along the way.''} (I9) While this study has the limitation of not having empirical evidence to suggest whether this approach is effective, it does align with the issue of social isolation identified by the instructors and with social learning theories \cite{vygotsky_mind_1978, bandura_social_1977}.

\subsection{Whose Responsibility Is It to Learn?}
When facing the dual pressure of learning harms and assessment harms from AI, even though instructors adopted different AI policies in practice, there is a fundamental shared understanding that continuing old practices as usual is not viable. When explaining the rationale behind their policies, the instructors converge to a core insight: The responsibility of ensuring students are learning is shifting.

\subsubsection{To Police or to Guide}
When discussing their own responsibilities, a significant majority of the instructors remarked on the infeasibility of actually enforcing AI-related rules in their classrooms (I1, I3, I4, I5, I9, I10, I11, I12, I13). This applies to instructors who put restrictions on AI use in their syllabus, such as for I13, who had a policy of submitting AI-generated code counts as academic integrity violation, but noted that \textit{``honestly, that's an empty threat. What are my highest priorities for improving my courses? Reporting students is a lower priority.''} It also lead to other instructors avoiding strict AI policies at all, including I9, who justified not putting restrictions on AI use because \textit{``I can only have a policy if I can enforce it. It's difficult to prove to a panel that something is LLM, even as an expert.''} 

As a result of the difficulty of enforcing AI policies, instructors began rethinking their roles in the class. A common refrain among the instructors is that they want to move away from the role of policing student behaviors or enforcing rules, and towards that of guiding students towards productive uses of AI (I1, I4, I8, I9, I10, I12). For example, I4 explains that:
\begin{quote}
    \textit{We've all started kind of crafting our policies and assignments so I don't have to play AI police officer, because I'm just not interested in that. I don't want to run around behind them and try and catch them using AI.}
\end{quote}
Beyond the toll it takes on instructors to try enforcing restrictive AI policies, some also believe that focusing on catching rule-breakers would create an adversarial relationship between instructors and students. For example, I8 noted: \textit{``I do believe in more rehabilitative or restorative justice, rather than just, like, isolation, punishment.''} Or they are concerned that it would harm the classroom culture (I1, I10), as I10 explains: \textit{``I have better things to do than chasing students about `is this AI-generated? Is this not AI-generated?' I also think it sort of creates an adverse learning environment, so I don't like that either.''} In contrast, instructors reported feeling better in the role of guiding students:
\begin{quote}
    \textit{I really like the shift of not policing the honor code part as much, and being like a guide and a coach, and providing information, but letting them learn on their own. I feel much better about that.} (I9)
\end{quote}
A more collaborative relationship between instructors and students could relieve pressure from instructors, and there are prior studies indicating that students prefer less adversarial AI policies as well \cite{sumilong_instructional_2025}. However, if, as I9 says, instructors could let students learn on their own, where does that leave the responsibility of ensuring students does not misuse AI? 

\subsubsection{Students are Responsible for Their Learning}
When instructors move away from policing their students' AI use, whether due to the intractability of detecting AI use or their teaching philosophy, it does not change the fact that many students will continue to use AI and some might become over-reliant. When instructors renounce some of their responsibilities in policing AI use, their AI policies naturally move towards having students be responsible for their AI use and its consequences. This can be reflected as an assessment policy, where some instructors place no restrictions on AI use or do not enforce AI restrictions, but would count on their assessments to show students who overuse AI the consequences (I2, I5, I7, I8, I12, I13). This offloading of responsibility reduces the workload on instructors, as I7 noted: \textit{`Because we've dropped the weight of the assignments so much, we quit looking for [cheaters]. [Proctored exams] sort of takes it off and puts it on the student to learn on their own.'} To other instructors, it also maintains the purpose of assessments as a means to sort their students based on their effort in learning:
\begin{quote}
    \textit{I'm now saying, yeah, go ahead and use it. The people who can only ask ChatGPT to generate something for them and turn it in are going nowhere. And I'm sort of okay with that, and maybe that's a little cynical for me, but it also means that \textbf{the students who are using it responsibly can be rewarded, and everyone else, well, too bad.}} (I2, emphasis ours)
\end{quote}
For instructors who rely less on assessments and focus more on learning, there is nevertheless a sense that students need to shoulder more responsibility in regulating their use of AI, as there is only so much instructors can tell them. For example I3 implemented AI policies that restricted some uses in the past, but later viewed it as a matter of student responsibility and dropped those policies:
\begin{quote}
    \textit{It's pretty obvious if the whole thing's copied from the GPT, and you do not even change the format. So I didn't really make any punishment on that\ldots We're all adults, and you should take responsibility for what you are doing. I mean, you are the person who's paying the tuition, you are the person who choose to take this course, choose to take this major. I just want you to learn.}
\end{quote}
Assigning responsibility of AI misuse to students can make sense, as students are assumed to have a degree of agency over the choice to use AI in their learning. When compounded with instructors having heavy workload and few effective means of detecting AI use, it is a logical step to count on students to self-regulate.

\subsubsection{But Should Students be Responsible?}
However, letting students bear the responsibility for learning harms entails diverting responsibility away from the AI tools and the companies behind them, who released these tools with few guardrails and numerous ethical concerns and technical issues like hallucination \cite{bender_dangers_2021}. While it is unlikely for instructors to hold AI tools or tech companies accountable, some instructors pointed out that it is not necessarily the students' fault for misusing AI either. As previously mentioned, I3 and I13 discussed how some AI tools can come pre-installed and impact students who did not actively intend to use them. However, I5 recognized that the pressure on students to use AI goes beyond the marketing of the tools themselves:
\begin{quote}
    \textit{AI is being injected everywhere, and \textbf{so many companies are just saying, `You better use AI while you're working.'} I don't know why that is the thing, but it is the command at Amazon and Microsoft from what we've heard from our graduates. They are forced to use AI, even if they may not even want to.} (emphasis ours)
\end{quote}
These external factors are layered on top of academic pressures students have always faced, such as time pressure. While study skills like time management and help-seeking are also commonly viewed as student responsibility, several instructors are reflecting upon whether they have the responsibility to alleviate pressure from students and thus steer them away from AI (I4, I6, I8, I9, I11):
\begin{quote}
    \textit{If I'm being charitable, and I want to be charitable with my students, I don't want to put them in a position where it's midnight, and the thing is due in the morning, and everyone's asleep, and I can't get anyone to user test this now, so I'll just use AI this one time. I think we end up putting our students in those positions. I mean, certainly they put themselves in that position, but I just don't want to make that position possible.} (I4)
\end{quote}
When viewed through the lens of instructor-student relationships, AI is not a new entity causing unprecedented problems. Rather, by creating new ways for students to learn (or not learn), it disrupts the prior equilibrium and forces instructors and students to renegotiate how they should teach and learn. While it is possible to offload some responsibilities to students due to practical concerns, I8 emphasized that the relationship between instructors and students should remain \textit{`reciprocal'}, and I4 found that the responsibility of instructors to students did not change fundamentally:
\begin{quote}
    \textit{Students feel guilty for using [AI] or, like, I'd look down on them or something\ldots I'm not judging them for any decisions they've made about AI or learning or anything like that. I want them to know that, hey, if you're ready to do well, I'm here to help you. Let's go. \textbf{I don't care what you've done before. I just want to help you. And I've had to express it in more concrete terms since AI.}} (emphasis ours)
\end{quote}

\section{Discussion}

\subsection{The Issues with Assessment-Oriented AI Policies}
As we have emphasized at the beginning, much of CS education researcher's attention has been dedicated to GenAI tools and how they can influence students. On the other hand, AI policies are completely different objects from AI tools. They are not digital technologies, but reified social contracts between students and instructors regarding AI technologies. They are made and implemented by CS instructors, yet exert direct influence on student learning experiences. It is thus crucial to avoid technological determinism, because the impact of AI cannot be understood by studying the capabilities of tools alone. In this study, we argue that understanding how CS instructors design and enforce AI policies, and how these policies mediate instructor-student relationships, are equally important to shaping the future of CS education.

Based on our reflexive thematic analysis of 13 semi-structured interviews with US higher education CS instructors, we find that AI policies are made both in response to AI's impact on student learning, and impact on instructors' ability to assess learning. Instructors have identified many ways GenAI tool misuse can harm student learning: AI impacts students cognitively, metacognitively, and socially; AI can hinder formation of basic skills, such as debugging, which influence their learning in the long term; and AI might have heterogeneous effect on different students mediated by their self-regulatory learning skills. Such complexity left many instructors uncertain about how they can address this first-order learning harm, especially when students are interacting with instructors less, also likely due to AI.

However, the interviews also surfaced assessment harms caused by AI, which is undermining the instructor's faith in their assessments, out of concern that students are submitting AI-generated work. The observation of some students performing well on take-home assignments but struggling in in-class exams led instructors to doubt whether many students have really learned. This distinct, second-order harm to the integrity of assessment and erosion of trust in students often led to higher-stake and more strict assessment policies, including filing large numbers of academic integrity cases, adopting proctored paper exams, and weighing summative assessments more heavily. These findings affirm prior research on how instructors planned to respond to AI \cite{lau_ban_2023}. This study provides new evidence on how instructors perceive the outcome of these policies: they acknowledge these assessment-oriented policies can be stress-inducing, less effective at evaluating learning, sometimes explicitly done with punitive intent, and ultimately indirectly addressing the GenAI tools at best. Nevertheless, assessment-oriented policies are often deemed the only way to counter the threat to assessment integrity.

However, AI policies that respond primarily to assessment harms exact second-order costs on instructor-student relationships. For one, instructors themselves found these policies not very cost-effective, as the labor of detecting AI is expensive. Even when students suffer negative consequences from AI misuse, they do not necessarily know more effective learning strategies. Disallowing AI misuse without corresponding positive intervention is akin to abstinence-only sex education, which has been demonstrated to be ineffective \cite{stanger-hall_abstinence-only_2011}. Assessment-oriented policies also put instructors in the role of policing students, creating adversarial classroom environments. Some instructors lamented the idea of being enforcers, and hope they can dedicate their effort to guiding students instead. 

The difficulty of evaluating student learning given AI also led some instructors to shift the responsibility of ensuring student learning to the student themselves. If there is no effective way for instructors to tell who used GenAI tools and intervene, the reasoning goes, it is up to students to regulate their own AI use. However, other instructors recognize that it is not fair to make students bear full responsibility, as GenAI tools are being aggressively pushed to students via pre-installation, marketing campaigns, and future career requirements. These factors, compounded with the always-present stressors of academic performance, led some instructors to question whether students should be responsible for their own learning struggles, or is AI more responsible and instructors should take a more proactive role in making policies to address AI directly.

\subsection{Toward Learning-Oriented AI Policies}
Fundamentally, AI is disrupting how instructors teach and how students learn. Ideally, students arrive in classrooms ready to learn, and instructors ready to teach. In practice, our performance-oriented way of structuring assessment and grading and the societal incentives in diplomas and job prospects dictate that students prioritize obtaining grades more than mastering knowledge and skills \cite{elliot_approach_1999}. Despite the sometimes competing incentives for students, by following pedagogical best practices, such as backward design \cite{wiggins_understanding_2005}, instructors can ensure learning and assessments are aligned such that the best way for students to obtain good grades is to master the content. It is thus unsurprising that AI proved so disruptive, since it provides a new, highly accessible way for students to pass assessments and obtain grades without necessarily engaging with the content. In response to this tipping of the balance on the student side, instructors can take two general approaches to restore equilibrium to the relationship---make assessments so punishing to students who over-relied on AI that they drive students back to learning, or make teaching more inclusive of AI such that students can recognize the option to learn with it. Given the difficulties instructors have experienced implementing the former, it is arguably the appropriate moment to examine the latter.

We should note that ``making teaching more inclusive of AI'' is neither a techno-solutionist plea for mass adoption of GenAI tools into teaching, nor a technological determinist judgment that AI will inevitably take over education and thus we must adapt. It is rooted in harm reduction \cite{marlatt_harm_1996}: We should recognize that students are using GenAI tools regardless of instructor's stances, and thus instructors should make AI policies that can best mitigate the learning harms associated with GenAI tools. Based on prior studies and the interviews in this study, there are several promising learning-oriented AI policy approaches that could receive more attention. 

\textbf{Transparency in AI Policies:} Prior research found that students dislike a lack of transparency in AI policies from instructors \cite{sumilong_instructional_2025, luo_jess_how_2025}, and several interviewed instructors likewise emphasized the importance of not only putting AI policies in writing, but clearly communicating the purpose and reasoning behind them. Without clarity of policy, students feel ashamed and nervous about disclosing their AI use, which makes it harder for instructors to provide support \cite{bao_ai_2025}. If instructors require student disclosure of their AI use, instructors should likewise be transparent about their own AI use.

\textbf{Model Good AI Use Cases:} There are a myriad of ways to use AI, from copy-pasting generated output to having in-depth conversations following the Socratic method. Crucially, there are legitimate reasons for each of these uses, given appropriate contexts. Penalizing students for over-reliance will not make students gain the ability to determine appropriate uses of AI, but explicitly teaching students AI literacy and modeling good uses could \cite{denny_prompt_2024, zamfirescu-pereira_why_2023}.

\textbf{Promote Metacognition with Formative Assessments:} While high-stake summative assessments that disallow AI could pressure students into not relying on AI, for students who are over-reliant, failing on the exam leaves them no chance to adjust their behavior. By providing students with lower-stake formative assessments, along with other metacognitive self-reflection activities, students could dispel their illusion of competence from AI use and convert to more self-regulated uses sooner \cite{prather_widening_2024, dawson_cognitive_2025}.

\textbf{Encourage Social Interactions and Inspire Motivation:} It is crucial to note that both prior work and instructors found AI eroding social learning in students as well \cite{hou_all_2025}. To counter AI becoming the default source of help, some instructors experimented with adding more open-ended and/or team-based projects, in hopes of students having stronger sense of ownership over their work and becoming more intrinsically motivated.

As we will discuss in the limitations, this study only involves interviews with instructors, so it does not have the empirical evidence to suggest whether these learning-oriented AI policies are effective and should be recommended. Still, given how assessment-oriented policies are causing frictions for both instructors and students, the learning-oriented approaches are promising sites for future research. CS education has an uncertain future due to the promises of GenAI, which can be both optimistic and grim. However, what matters at least as much is CS education in the here and now, and our study presents worrying findings on how GenAI frayed instructor-student relationships, drove instructors toward assessment policies that are less about supporting student learning, and left students bearing more responsibility for their own behaviors. Given these challenges instructors and students are facing right now, researchers and practitioners alike should focus on developing AI policies that can restore trust with students and guide them toward constructive and discerning uses of AI.

\subsection{Limitations and Future Work}
While we strived to recruit instructors that teach in a wide range of learning environments, in an effort to define the scope of this study, we recruited instructors from US institutions only. As such, the findings should be contextualized within this limited scope, as attitudes towards GenAI in education and policies made in response could both vary across countries. Similarly, this interview study focuses on undergraduate CS education, and thus any interpretations of the data is limited to that context. Dynamics surrounding AI policy in K-12 or post-graduate environments would require further research. The study focuses on the relationship between students and instructors, and thus relationships that are also relevant to the regulation of student AI use, such as interactions between instructors and academic integrity offices, are outside the scope of this study.

The interview protocol asks instructors about their observations of students, including how student performance has changed since the introduction of GenAI tools, and what effects their AI policies have had on students. However, the answers to these questions are solely self-reported accounts by the instructors, without another source of empirical data to validate them. The findings should therefore be interpreted as what the instructors perceive from their personal experiences, but need to be verified by future work.

While this study focuses on learning-oriented policies that are relatively easy to implement, CS education might still need to adjust learning objectives and learning design in the long term. Given the capability of AI to automate basic coding tasks, skills like memorizing syntax might be less necessary for some students, such as 

\section{Conclusion}
In this study, we conducted 13 interviews with undergraduate CS instructors to understand how and why they design and implement AI policies in their courses. Complementing the existing research on how students and instructors view and use GenAI tools, our findings reveal how instructor-student relationships are indirectly influenced by AI. AI policies are reified representations of how instructors are rethinking their roles and responsibilities to students in light of AI. The common AI policies that focus on AI-proofing assessments could deter student AI use in the short term, but can place additional stress on students and strain instructor-student relationships in the long term. While mitigating AI's learning harms on students is difficult, we advocate for instructors to shift focus away from AI-proofing assessments and become more transparent and inclusive to student AI use.

\bibliographystyle{ACM-Reference-Format}
\bibliography{sample-base}

\end{document}